\documentclass[floats,floatfix,showpacs,amssymb,prd,superscriptaddress,twocolumn,aps,nofootinbib]{revtex4-1}
\usepackage{graphicx, epsfig, amssymb} 
\usepackage{amsmath, amsfonts}
\usepackage{bm} 
\usepackage[breaklinks]{hyperref}
\usepackage{color}

\def\nn{\nonumber}
\def\be{\begin{equation}}
\def\ee{\end{equation}}
\def\beq{\begin{eqnarray}}
\def\eeq{\end{eqnarray}}

\newcommand{\ADM}{{\mbox{\tiny ADM}}}

\begin{document}

\title{On generic parametrizations of spinning black-hole geometries}

\author{Vitor Cardoso}
\affiliation{CENTRA, Departamento de F\'{\i}sica, Instituto Superior T\'ecnico, Universidade de Lisboa,
Avenida Rovisco Pais 1, 1049 Lisboa, Portugal.}
\affiliation{Perimeter Institute for Theoretical Physics, Waterloo, Ontario N2L 2Y5, Canada.}
%

\author{Paolo Pani}
\affiliation{CENTRA, Departamento de F\'{\i}sica, Instituto Superior T\'ecnico, Universidade de Lisboa,
Avenida Rovisco Pais 1, 1049 Lisboa, Portugal.}
\affiliation{Institute for Theory and Computation, Harvard-Smithsonian
CfA, 60 Garden Street, Cambridge, MA, USA.}

\author{Jo\~ao Rico}
\affiliation{CENTRA, Departamento de F\'{\i}sica, Instituto Superior T\'ecnico, Universidade de Lisboa,
Avenida Rovisco Pais 1, 1049 Lisboa, Portugal.}

\begin{abstract} 
The construction of a generic parametrization of spinning geometries which can be matched continuously
to the Kerr metric is an important open problem in General Relativity. Its resolution is of more than academic interest, as it allows to parametrize and quantify possible deviations from the no-hair theorem.
Various approaches to the problem have been proposed, all with their own (severe) limitations. Here we discuss the metric recently proposed by Johannsen and Psaltis, showing that (i) the original metric describes only corrections that preserve the horizon area-mass relation of nonspinning geometries; (ii) this unnecessary restriction can be relaxed by introducing a new parameter that in fact dominates in both the weak-field and strong-field regimes; (iii) within this framework, we construct the most generic spinning black-hole geometry which contains twice as many (infinite) parameters as the original metric; (iv) in the strong-field regime, all parameters are (roughly) equally important. This fact introduces a severe degeneracy problem in the case of highly-spinning black holes. Our results suggest that using parametrizations that affect only the quadrupole moment of the Kerr geometry is problematic, because higher-order multipoles can be equally relevant for highly-spinning objects. Finally, we prove that even our generalization fails to describe the few known spinning black-hole metrics in modified gravity.
\end{abstract}

\pacs{04.70.Bw, 04.50.Kd, 04.70.-s}

\maketitle

\section{Introduction}
From the astrophysical viewpoint, one of the most important predictions of General Relativity (GR) is that isolated spinning black holes (BHs) are universally described by the Kerr family. The latter is uniquely defined by only two parameters, the mass ${M_\ADM}$ and the angular momentum $J$ (see e.g. Ref.~\cite{Wiltshire:2009zza}). Tests of GR in the strong-field regime hinge on validation of this ``Kerr hypothesis.''
Such tests will become possible in the near future thanks to novel electromagnetic~\cite{Eisenhauer:2008tg,2013arXiv1311.5564B} and gravitational-wave~\cite{Berti:2009kk,Yunes:2013dva} observations. It is thus of utmost importance to develop theoretical tools that allow to test the nature of massive compact objects in the near-horizon region.

These two --~electromagnetic and gravitational~-- windows are, in practice, probes to different dynamical regimes. Gravitational waves probe highly-dynamical configurations and therefore the full content of the field equations, while electromagnetic observations typically involve the motion of ``test particles'' in a fixed background geometry. This distinction is important because large classes of theories have the Kerr metric as a solution~\cite{Psaltis:2007cw},
and would therefore be indistinguishable in the electromagnetic window.
In general however, any extension of GR would affect BH solutions, for example by modifying the multipolar structure~\cite{Geroch:1970cd,Hansen:1974zz} of spinning BH spacetimes and their near-horizon geometry~\cite{Vigeland:2009pr,Vigeland:2010xe}. Thus, a case-by-case analysis is quixotic and much effort has been recently devoted to construct model-independent spinning geometries that can be used as alternatives to the Kerr metric.

Several approaches have been proposed, each of them with their own limitations (cf. Ref.~\cite{Joao} for a discussion). For example, the original bumpy BH formalism~\cite{Collins:2004ex,Vigeland:2009pr} assumes Einstein's equations, whereas the ``quasi-Kerr'' spacetimes~\cite{Glampedakis:2005cf} are not regular close to the horizon. In the modified bumpy BH formalism~\cite{Vigeland:2011ji}, spinning geometries are regular~\cite{Johannsen:2013rqa}, but their construction assumes the existence of an approximate Carter constant.

To overcome these limitations, Johannsen and Psaltis (JP) have recently proposed a novel approach~\cite{Johannsen:2011dh}, where a modified Kerr geometry is obtained by applying the Newman-Janis algorithm~\cite{NJ} to a deformed Schwarzschild metric. At variance with previous studies, this approach does not assume Einstein's equations, nor the existence of an approximate Carter constant. Even though the procedure makes use of the --unjustified, because the field equations are unknown-- Newman-Janis transformation (see e.g. Ref.~\cite{Hansen:2013owa} for some criticism),
the final transformed metric could as well be the ad-hoc starting point for the investigation of deviations from GR~\cite{Joao}.
The JP metric has been widely used for tests involving observations of the images of inner accretion flows, X-ray observations of relativistically broadened iron lines or of the continuum spectra of accretion disks (cf. e.g.~\cite{Johannsen:2011dh,Bambi:2011ek,Johannsen:2013rqa,Johannsen:2013asa} and references therein). This metric has been found to contain naked singularities~\cite{Johannsen:2013rqa} (see also~\cite{Bambi:2011ew,Bambi:2011yz}). However, such singularities only appear near the extremal limit for sufficiently small deformations, and they correspond to the fact that the maximum spin of the near-extremal deformed geometry is smaller for some choice of the parameters~\cite{Johannsen:2013rqa}.


The scope of this paper is to extend the parametrization proposed in Ref.~\cite{Johannsen:2011dh} in various ways and to discuss some limitations of this approach.

\section{Analysis of the JP metric}
The seed of the JP metric is the static and spherically symmetric line element (we use $G=c=1$ units):
\begin{equation}
ds^2=-f[1+\bar{h}(r)]dt^2+f^{-1}[1+\bar{h}(r)]dr^2+r^2d\Omega^2\,, \label{static1}
\end{equation}
where $d\Omega^2=d\theta^2+\sin^2\theta d\phi^2$, $f\equiv 1-2M/r$ and 
\begin{equation}
\bar{h}(r)\equiv\sum_{k=0}^\infty \epsilon_k\left({M}/{r} \right)^k\,. \label{eq:hr_JP}
\end{equation}
By applying the Newman-Janis algorithm to the metric~\eqref{static1}, one arrives at the modified Kerr metric in Boyer-Lindquist coordinates~\cite{Johannsen:2011dh}
\begin{eqnarray}
ds^2&&=-[1+h]\left(1-\frac{2Mr}{\Sigma}\right)dt^2-\frac{4aMr\sin^2\theta}{\Sigma}[1+h]dtd\phi\nn\\
&&+\frac{\Sigma[1+h]}{\Delta+a^2\sin^2\theta h}dr^2+\Sigma d\theta^2 +\left[ r^2+a^2\right.\nonumber \\
&&\left.+\frac{2a^2Mr\sin^2\theta}{\Sigma}+h\frac{a^2(\Sigma+2Mr)\sin^2\theta}{\Sigma} \right]\sin^2\theta d\phi^2\,, \label{metricJP}
\end{eqnarray}
where $\Sigma\equiv r^2+a^2\cos^2\theta,\,\Delta\equiv r^2-2Mr+a^2$ and $h=h(r,\theta)$ is related to the original $\bar{h}(r)$ through the Newman-Janis transformation, and reads
\begin{equation}
h(r,\theta)\equiv\sum^\infty_{k=0}\left(\epsilon_{2k}+\epsilon_{2k+1}\frac{Mr}{\Sigma}\right)\left(\frac{M^2}{\Sigma}\right)^k\,. \label{hJP}
\end{equation}
In Ref.~\cite{Johannsen:2011dh} it was argued that asymptotic flatness requires $\epsilon_0=\epsilon_1=0$ and that observational bounds coming from Lunar Laser Ranging experiments~\cite{Williams:2004qba} translate in the constraint $|\epsilon_2|\leq 4.6\times 10^{-4}$.
In the simplest incarnation of the metric, all parameters $\epsilon_n$ are set to zero, except for the first (in a weak-field expansion) unconstrained parameter $\epsilon_3$. The function~\eqref{hJP} therefore is now given by $h(r,\theta)=\epsilon_3{M^3r}/{\Sigma^2}$.

The properties of this simplified metric have been studied by several authors (cf. e.g.~\cite{Johannsen:2011dh,Bambi:2011ek,Johannsen:2013rqa,Johannsen:2013asa} and references therein), including the shape of the event horizons, the existence of naked singularities in a certain region of the parameter space, and the structure the inner disc edge instabilities as a function of the extra parameter $\epsilon_3$ and of the spin $a/M$. The degeneracy between $a/M$ and $\epsilon_3$ in the X-ray emission from accretion discs has been investigated in several studies (see e.g.~\cite{Johannsen:2012ng,Bambi:2013eb} and references therein). 

In the following, we point out three important facts that have seemingly been overlooked in all studies subsequent to Ref.~\cite{Johannsen:2011dh}: 

\noindent i) The condition $\epsilon_1=0$ is superfluous even in the asymptotically flat case and, in fact, the corrections associated to the parameter $\epsilon_1$ are dominant in the entire region of the parameter space;

\noindent ii) The choice of $\bar h(r)$ is not unique and, in particular, Eq.~\eqref{eq:hr_JP} is a weak-field expansion that is valid at $r\gg M$. For highly-spinning geometries, quantities like the innermost stable circular orbit (ISCO) are closer to the horizon and all parameters $\epsilon_k$ are roughly equally important. This fact introduces a severe degeneracy problem in the case of highly-spinning BHs, specially when only a few parameters in the series~\eqref{hJP} are considered; 

\noindent iii) Finally, even without imposing any restriction on $\epsilon_1$, the metric~\eqref{metricJP} can be further extended,  as shown in Sec.~\ref{sec:generalization}.

The rest of the paper is devoted to derive these results and to discuss some further limitations of this approach.

\section{Generalizations of the JP metric}
\subsection{Dominant correction}
Let us first show that the condition $\epsilon_1=0$ imposed in Ref.~\cite{Johannsen:2011dh} and followed in all subsequent work is not necessary to ensure asymptotic flatness. Indeed,
imposing only $\epsilon_0=0$, the asymptotic expansion of the metric~\eqref{metricJP} reads
\begin{eqnarray}
-g_{tt}&\to&1-\frac{2{M_\ADM}}{r}+\frac{{M^2_\ADM}}{r^2}\frac{(\epsilon_2-2\epsilon_1)}{\left(1-\epsilon_1/2\right)^2}+{\cal O}(1/r^3)\,, \label{gttinf}\\
g_{rr}&\to &1+\left(\frac{2+\epsilon_1}{2-\epsilon_1}\right)\frac{2M_\ADM}{r}+{\cal O}(1/r^2)\,,\\
g_{tr}&\to& -\frac{2J}{r}\sin^2\theta+{\cal O}(1/r^2) \label{gtrinf} \,,
\end{eqnarray}
where ${M_\ADM}$ is Arnowitt-Deser-Misner mass of the solution~\eqref{metricJP} and $J$ is the BH angular momentum. These quantities are related to the parameters $M$ and $a$ appearing in the metric through
\begin{eqnarray}
 {M_\ADM}&=&M\left(1-{\epsilon_1}/{2}\right)\,,\label{mass}\\
 J&\equiv& a M =\frac{a{M_\ADM}}{1-{\epsilon_1}/{2}}  \,.\label{spin}
\end{eqnarray}
It is also easy to show that the quadrupole moment of the solution is affected by $\epsilon_1$, $\epsilon_2$ and $\epsilon_3$.

Equation~\eqref{mass} shows a simple property that was ignored in previous analysis of the metric~\eqref{metricJP}: the parameter $\epsilon_1$ is associated to a mass shift with respect to the Schwarzschild solution. It is easy to show that the BH area ${\cal A}$ normalized by the total mass ${M_\ADM}$ in the nonrotating case reads
\begin{equation}
 \frac{{\cal A}}{16\pi{M^2_\ADM}}=\frac{1}{(1+\epsilon_1/2)^2}\sim1+\epsilon_1+{\cal O}(\epsilon_1^2)\,, \label{area}
\end{equation}
where in the last step we have expanded the results to first order in the small-$\epsilon_1$ limit. In this case the horizon of the nonrotating metric~\eqref{static1} is simply located at $r_H=2M$. Thus, the horizon area is affected only by the first nontrivial parameter, $\epsilon_1$.
In other words, imposing $\epsilon_1\equiv0$ restricts the space of solutions described by Eq.~\eqref{metricJP} to those whose horizon area in the nonrotating case coincides with that of the undeformed Schwarzschild BH with the same mass and in the same coordinates.

However, the deformations associated to $\epsilon_1$ and $\epsilon_2$ are strongly constrained by weak-field tests. Indeed, by matching Eqs.~\eqref{gttinf}--\eqref{gtrinf} with a parametrized post-Newtonian (PPN) expansion~\cite{Will:2005va},
\begin{eqnarray}
 -g_{tt}&\to&1-\frac{2{M_\ADM}}{r}+2(\beta-\gamma)\frac{{M^2_\ADM}}{r^2}+{\cal O}(1/r^3)\,, \label{gttPPN}\\
  g_{rr}&\to &1+2\gamma\frac{M_\ADM}{r}+{\cal O}(1/r^2) \,, \label{grrPPN}
\end{eqnarray}
we can easily identify $\epsilon_1=2-4/(1+\gamma)$ and $\epsilon_2={4 [2 \beta-1 +(\gamma-2 ) \gamma]/(1+\gamma )^2}$. The PPN parameters are very well constrained by observations and their measured value is close to unity\footnote{Note however that the PPN constraints are derived assuming the central object is a star so, in order to apply these bounds, we have to make the extra assumption that the asymptotic behaviors~\eqref{gttinf}--\eqref{gtrinf} are the same for a BH geometry and a star. This might not be the case in some modified gravity, for example in theories which allow for some Vainshtein-like mechanism, cf. Ref.~\cite{environment} for a discussion.}, $|\gamma-1|\lesssim10^{-5}$ and $|\beta-1|\lesssim 2.3\times10^{-4}$~\cite{Will:2005va,Williams:2004qba}. These constraints translate in $|\epsilon_1|\lesssim 10^{-5}$ and $|\epsilon_2|\lesssim4.6\times 10^{-4}$.

Note that the changes introduced by $\epsilon_1$ are physical and cannot be absorbed in some coordinate transformation. For example, one could normalize all lengths by the total mass ${M_\ADM}$ and one would find that the horizon location (in the Schwarzschild coordinates used in Eq.~\eqref{static1}) is different from the Schwarzschild case, $r_H/{M_\ADM}\sim 2+\epsilon_1$ in the small-$\epsilon_1$ limit.

The fact that the horizon area-mass ratio is different from Schwarzschild is a very natural property of BH solutions in modified gravity, as the near-horizon geometry and the total mass of the spacetime are affected by any putative extra field. This is the case, for example, in theories that allow for BHs with secondary hair (cf. e.g. Refs.~\cite{Pani:2009wy,Barausse:2011pu}).

\begin{figure}[t]
\begin{center}
\begin{tabular}{c}
\epsfig{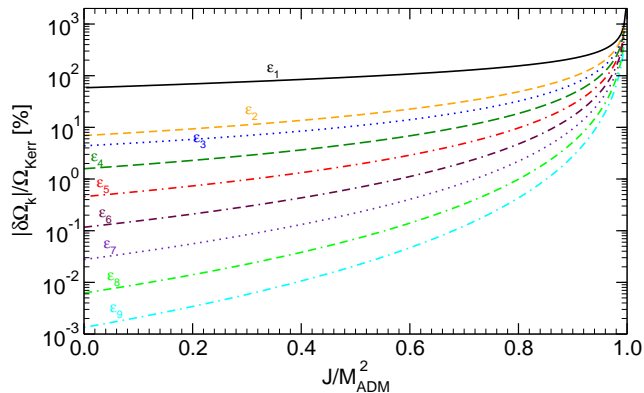}
\end{tabular}
\caption{Relative corrections $\delta\Omega_k/\Omega_0$ to the ISCO frequency as a function of $J/{M^2_\ADM}$ for the metric~\eqref{metricJP} to linear order in $\epsilon_k\ll1$ up to $k=9$. The ISCO frequency reads $\Omega=\Omega_0+\sum_k\delta \Omega_k\epsilon_k$, where $\Omega_0$ is the ISCO frequency of a Kerr geometry. The small-coupling approximation requires $(\delta \Omega_k/\Omega_0)\epsilon_k\ll1$ for consistency.
Each $\epsilon_k-$line is built by setting to zero all other $\epsilon_i,\,i\neq k$.
\label{fig:deltaOmegaISCO}}
\end{center}
\end{figure}
%

\subsection{Curvature singularities}\label{sec:naked0}
It has been shown that the simplest incarnation of the original JP metric~\eqref{metricJP} (i.e. setting $\epsilon_3$ as the only nonvanishing extra parameter) possesses naked singularities. However, when $\epsilon_3$ is small such singularities appear only close to the limit $J\to M_\ADM^2$, like in the Kerr geometry~\cite{Johannsen:2013rqa}. 
It is therefore relevant to understand if this picture changes when more coefficients $\epsilon_k$ are considered.

In order to gain insight on this issue, it is convenient to work within a perturbative scheme in $\epsilon_k$. 
The motivation to work perturbatively in powers of $\epsilon_k$ is twofold. On the one hand the perturbative expansion simplifies the analysis, without the need of resorting to numerics. On the other hand this expansion is motivated by the fact that the coefficients $\epsilon_k$ should be thought as related to the extra parameters appearing in a putative modified theory of gravity. The latter can be considered as an effective field theory which is valid only perturbatively in the small-coupling limit, i.e. only to some order in $\epsilon_k\ll1$.
With this motivation, in this section we extend the analysis of Ref.~\cite{Johannsen:2013rqa} by turning on more parameters $\epsilon_k$ and working to ${\cal O}(\epsilon_k)$. 

In such perturbative expansion, the metric~\eqref{metricJP} is a weak-field deformation of the Kerr solution and it is clear that possible naked singularities might only appear in the limit $J\to M_\ADM^2$, i.e. when the unperturbed metric is marginally regular (cf. Fig.~3 in Ref.~\cite{Johannsen:2013rqa}).

First of all, let us compute the event horizon. For generic stationary and axisymmetric spacetimes, the latter is defined as the locus $r_+(\theta)$ which satisfies the ordinary differential equation~\cite{Johannsen:2013rqa}
\begin{equation}
 g^{rr}-2 g^{r\theta} \frac{dr_+(\theta)}{d\theta}+g^{\theta\theta}\left(\frac{dr_+(\theta)}{d\theta}\right)^2=0\,,
\end{equation}
where the metric coefficients are evaluated at $r=r_+(\theta)$.
Because in the coordinates used in Eqs.~\eqref{eq:JP2_1}--\eqref{eq:JP2_5} the cross term $g^{r\theta}=0$, and because $dr_+(\theta)/d\theta=0$ to zeroth order in $\epsilon_k$, to first order the equation that defines the horizon location is algebraic and simply reads $g^{rr}=0$ as in GR. It is then straightforward to compute the horizon location to ${\cal O}(\epsilon_k)$ for any $k$. To make our point, it is sufficient to consider $\epsilon_1=\epsilon_2=0$ and $\epsilon_k=0$ when $k>4$, i.e., we only take $\epsilon_3$ and $\epsilon_4$ to be nonvanishing. The result reads
\begin{eqnarray}
 r_+(\theta)&&=\frac{D}{{M_\ADM}}+{M_\ADM}\nn\\
 &&+\frac{ {\sin^2\theta}{M_\ADM^5} \left(D-{M_\ADM^2}\right)  \left(D \epsilon_3+{M_\ADM^2} (\epsilon_3+\epsilon_4)\right)}{2 D \left(D+{M_\ADM^2}\right) \left(2 {M_\ADM^2}+\left(D-{M_\ADM^2}\right) {\sin^2\theta}\right)^2}\nn\\
 &&+{\cal O}(\epsilon_k^2)\,,\label{rhJP}
\end{eqnarray}
where we have defined $D^2=M_\ADM^4-J^2$ and the extremal Kerr limit corresponds to $D\to0$. The result above reduce to the perturbative analysis done in Ref.~\cite{Johannsen:2013rqa} when $\epsilon_4=0$. 

It is easy to show that the ${\cal O}(\epsilon_k)$ correction to the horizon is regular everywhere except when $D\to0$. In this limit, the divergence of the horizon location is then reflected on curvature invariants such as the Kretschmann scalar $R_{abcd}R^{abcd}$ at the horizon. 


Clearly, when the linear correction diverges the perturbative expansion breaks down. However, it can be easily verified numerically that a curvature singularity generically exists beyond the linear level, as it was explicitly shown in Ref.~\cite{Johannsen:2013rqa}. Nonetheless, the existence of these singularities simply restrict the parameter space to the region $J<J_{\rm crit}(\epsilon_k)$. As shown above $J_{\rm crit}\to M_\ADM^2$ when the deformations are small.

Interestingly, from Eq.~\eqref{rhJP} one can see that when $\epsilon_4=-\epsilon_3$ the geometry is everywhere regular at $r\geq r_+(\theta)$ even when $J=M_\ADM^2$, precisely like the Kerr metric. This fact was missed in the analysis of Ref.~\cite{Johannsen:2013rqa} because $\epsilon_4=0$ in that case. Note that the same result holds for higher-order coefficients: curvature singularities at $J=M_\ADM^2$ can always be avoided if $\epsilon_6=-\epsilon_5$, $\epsilon_8=-\epsilon_7$ and so on. 

Although a detailed study of the regularity of the metric beyond the linear level would be interesting, our results already show that a generalization of the simplest JP metric can be everywhere regular at least to first order in an effective-field theory approach and in the same region of the mass-spin parameter space in which the Kerr metric is regular, i.e. $J\leq M_\ADM^2$.

\subsection{Dependence on higher-order parameters}
Leaving aside possible observational constraints and the regularity of the geometry, in this section we explore the relative relevance of the various parameters $\epsilon_k$ in the JP metric~\eqref{metricJP}. As an example, we compute the ISCO frequency for different combinations of nonvanishing $\epsilon_k$ and for different values of the BH spin. 

For simplicity, we assume that the underlying modified theory of gravity satisfies the weak equivalence principle, i.e. we assume that test particles move along geodesics of the spacetime. This assumption can be easily relaxed (cf. e.g. Ref.~\cite{Pani:2009wy}) and it is not crucial not show our point.
The geodesic motion of a stationary and axisymmetric spacetime is governed by the effective potential
\begin{equation}
V_{\rm eff}(r,\theta)=\frac{g_{\phi\phi}E^2+2g_{t\phi}E L+g_{tt}L^2}{g^2_{t\phi}-g_{tt}g_{\phi\phi}}-1\,, \label{eq:efpot2}
\end{equation}
where the energy and angular momentum respectively reads $E=-g_{tt} \dot{t}-g_{t\phi}\dot{\phi}$ and $L_z=g_{t\phi}\dot{t}+g_{\phi\phi}\dot{\phi}$, and they are constants of motion [here and in the following a dot and a prime denote time and radial derivatives, respectively]. Circular equatorial orbits are defined by $\theta=\pi/2={\it const}$ and $V_{\rm eff}(r_c,\pi/2)=V_{\rm eff}'(r_c,\pi/2)=0$. For a given circular orbit, the orbital frequency reads
\begin{equation}
\Omega(r_c)=\left.\frac{-g_{t\phi}'\pm\sqrt{g_{t\phi}'^2-g_{tt}' g_{\phi\phi}'}}{g_{\phi\phi}'}\right|_{r=r_c}. \label{omega2}
\end{equation}
Finally, the ISCO location $r_{\rm ISCO}$ is defined by the further condition $V_{\rm eff}''(r_{\rm ISCO},\pi/2)=0$ and the ISCO frequency reads $\Omega(r_{\rm ISCO})$. 
Note that, in principle, stability of circular orbits should also be studied under vertical perturbations. However, the vertical stability of the ISCO of a Kerr BH guarantees that the ISCO of the deformed metric is also stable if the parameters $\epsilon_k$ are small enough, i.e. $\partial^2_\theta V_{\rm eff}(r_{\rm ISCO},\pi/2)<0$ for any $J$ in the small-$\epsilon_k$ limit. Away from this limit, the deformed ISCO can develop a vertical instability depending on the values of $J/M_\ADM^2$ and of $\epsilon_k$, cf. Ref.~\cite{Joao} and references therein.

Implementing the equations above for the metric~\eqref{metricJP} is straightforward. By allowing more parameters $\epsilon_k$ to be nonvanishing, it is possible to show that there exists a degeneracy problem: for any value of the spin, the gauge-invariant ISCO frequency can take its Kerr value for multiple combinations of the parameters $\epsilon_k$. Thus, any experimental or observational approach based on the assumption that $\epsilon_k=0$ for $k\geq 4$ is rendered blind to the possibility that the observed spacetime, for which a certain $\epsilon_3$ and spin $J$ would be estimated, is in reality a spacetime with, say, a much larger value of $\epsilon_3$ and a non-zero value of $\epsilon_4$. This is a fundamental difference from, e.g., Ryan's approach~\cite{Ryan:1995wh} where the estimates of each of the multipole moments, such as the mass, spin, and quadrupole, are independent of the higher-order moments. 

A perturbative approach is perhaps more illuminating. In a small-$\epsilon_k$ expansion any observable (e.g. the ISCO frequency) can be written as
\begin{equation}
 X=X_{\rm{Kerr}}+\sum_{k=1}^\infty \delta X_{k}\epsilon_k+{\cal O}(\epsilon_k^2)\,,\label{eq:small_eps_exp}
\end{equation}
where $X_{\rm{Kerr}}$ is the value of the quantity $X$ for the Kerr metric. The corrections $\delta \Omega_k$ to the orbital frequency can be computed analytically, but their exact form is of no interest here. Such corrections are shown in Fig.~\ref{fig:deltaOmegaISCO} as a function of the spin for $X$ equal to the ISCO frequency and normalized by the value $\Omega_{\rm Kerr}$ of a Kerr BH with same mass ${M_\ADM}$ and same spin $J$.

For relatively small values of the spin, we observe a hierarchy between different parameters: the higher the order of $k$, the smaller the correction $\delta \Omega_k$. Since Eq.~\eqref{eq:hr_JP} is a \emph{far field} expansion, it is not guaranteed in principle that it will converge in the strong-field region near the ISCO. Indeed, such hierarchy deteriorates in the near-extremal limit, $a\to M$, because the ISCO location is close to the horizon and stable orbits can probe regions of stronger curvature \footnote{A similar argument applies also to gravitational-wave tests using background geometries with weak-field parametrizations~\cite{environment}. This is due to the fact that geodesic motion has a lower cutoff in frequency given by the ISCO, whereas the ringdown emission is governed by the light ring~\cite{Berti:2009kk}. In the slowly-rotating case, both the ISCO and the light ring radii are larger than the central mass $M_{\ADM}$, so higher powers of $M_{\ADM}/r$ are suppressed. However, in the highly-spinning case both cutoffs are comparable to the BH mass $M_{\ADM}$, so the convergence of the weak-field expansion deteriorates.}. 

As shown in Fig.~\ref{fig:deltaOmegaISCO}, all linear corrections are roughly equally important for highly-spinning BHs.
For example, when $a=0.998M$, $|\delta\Omega_k|/\delta\Omega_1\sim(0.70,0.65,0.59,0.54)$ for $k=(7,8,9,10)$, respectively.
Qualitatively similar results hold true also in the exact case, i.e. away from the small-coupling approximation. In that case, it is easy to show that the relative difference between the corrections associated to two generic coefficients $\epsilon_{k_2}$ and $\epsilon_{k_1}$ (with $k_2>k_1$) decreases for large spin, showing that higher order corrections become relatively more important.

\subsection{A further generalization}\label{sec:generalization}
We now study a further generalization of the metric~\eqref{metricJP} and briefly explore its properties. Our starting point is to recognize that the static metric~\eqref{static1} is restricted by the unnecessary requirement that the perturbations of the $g_{tt}$ and $g_{rr}$ components are the same. This property imposes too strong a restriction on the form of the (effective) stress-energy tensor that modifies Einstein's equations.
Examples of theories where this restriction does not hold are studied for instance in Refs.~\cite{Mignemi:1992nt,Pani:2011gy,Barausse:2011pu} and, generically, static solutions in modified theories will not be described by a metric in the form of Eq.~\eqref{static1}. We then consider the seed metric
\begin{equation}
ds^2=-f(1+\bar h^t) dt^2+f^{-1}(1+\bar h^r)dr^2+r^2d\Omega^2\,,\label{static2}
\end{equation}
which represents the most general spherically symmetric geometry with no further restrictions (see e.g. Ref.~\cite{Chandra}). Following the original prescription~\cite{Johannsen:2011dh}, we expand the functions $\bar h^t$ and $\bar h^r$ in powers of $M/r$ as
\begin{equation}
\bar h^i(r)\equiv \sum\limits_{k=0}^{\infty} \epsilon^i_k \left(\frac{M}{r}\right)^k, \qquad i=t,r.
\end{equation}
By applying the standard Newman-Janis algorithm, we obtain the generalized deformed Kerr metric~\cite{Joao}
\begin{eqnarray}
g_{tt}&=&-F(1+h^t), \label{eq:JP2_1}\\
g_{rr}&=&\frac{\Sigma (1+h^r)}{\Delta+a^2 \sin^2\theta h^r},\label{eq:JP2_2}\\
g_{\theta\theta}&=&\Sigma,\label{eq:JP2_3}\\
g_{\phi\phi}&=&\sin^2\theta \left\{\Sigma + a^2 \sin^2\theta \left[2 H - F (1+h^t) \right] \right\},\label{eq:JP2_4}\\
g_{t\phi}&=&-a\sin^2\theta \left[H-  F (1+h^t)\right], \label{eq:JP2_5}
\end{eqnarray}
where we have introduced $F\equiv 1-2Mr/\Sigma$, $H\equiv\sqrt{(1+h^r)(1+h^t)}$, and $h^i=h^i(r,\theta)$ are the generalizations of Eq.~\eqref{hJP} with the substitution $\epsilon_k\to\epsilon_k^i$ for $i=t,r$. Note that this parametrization reduces to the original one, Eq.~\eqref{metricJP}, when $\epsilon_k^t\equiv\epsilon_k^r$ for any $k$, and it contains twice as many (infinite) parameters as those entering in the original metric~\eqref{metricJP}.

\begin{figure}[t]
\begin{center}
\begin{tabular}{c}
\epsfig{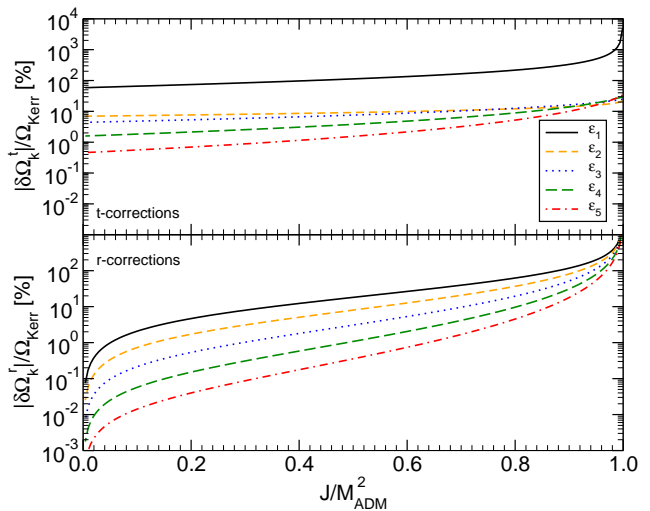}
\end{tabular}
\caption{Same as Fig.~\ref{fig:deltaOmegaISCO} for the generalized metric~\eqref{eq:JP2_1}--\eqref{eq:JP2_5}. The two panels refer to the corrections associated to $\epsilon_k^t$ (upper panel) and $\epsilon_k^r$ (lower panel), respectively. For ease of comparison, the range of the vertical axis is the same for both panels. In this case the total ISCO frequency reads $\Omega=\Omega_0+\sum_k\delta \Omega_k^t\epsilon_k^t+\sum_k\delta \Omega_k^r\epsilon_k^r$. The small-coupling approximation requires $(\delta \Omega_k^i/\Omega_0)\epsilon_k^i\ll1$ for consistency.
\label{fig:deltaOmegaISCO2}}
\end{center}
\end{figure}

Imposing asymptotic flatness requires again only $\epsilon_0^t=\epsilon_0^r=0$, but does not impose any constraint on $\epsilon_1^t$ and $\epsilon_1^r$, similarly to the case previously discussed. Expanding the metric elements~\eqref{eq:JP2_1} and \eqref{eq:JP2_2} at infinity and comparing with the PPN expansions~\eqref{gttPPN} and \eqref{grrPPN}, we can identify 
\begin{eqnarray}
  {M_\ADM}&=&M\left(1-{\epsilon_1^t}/{2}\right)\,,\label{mass2}\\
  \epsilon_1^r&=&-2-\gamma(\epsilon_1^t-2)\,, \label{constr1}\\
  2\epsilon_2^t&=&(\beta-\gamma)(\epsilon_1^t-2)^2+4\epsilon_1^t \label{constr2}\,.
\end{eqnarray}
Therefore, even imposing the GR values $\beta=\gamma=1$ supported by observations, the parameter $\epsilon_1^t$, as well as $\epsilon_2^r$ and all parameters $\epsilon_k^i$ with $k>2$ and $i=t,r$, are left unconstrained.

Figure~\ref{fig:deltaOmegaISCO2} shows the shifts of the ISCO frequency for the generalized metric~\eqref{eq:JP2_1}--\eqref{eq:JP2_5} in the small-$\epsilon_k^i$ limit, obtained using an expansion analogous to Eq.~\eqref{eq:small_eps_exp}. 
For low rotation rates the corrections associated to $\epsilon_k^t$ are larger than those associated to $\epsilon_k^r$, while the reverse is true for the highly-spinning case, $a/M\gtrsim 0.85$. An exception to this behavior are the $\epsilon_1^i$ parameters in which the $t$--correction is larger than the $r$--correction for any spin. The fact that the $\epsilon_k^r$ corrections are negligible in the small-spin limit stems from the fact that the effective potential~\eqref{eq:efpot2} does not involve the $g_{rr}$ component, which is the only one containing the $\epsilon_k^r$ terms in the nonrotating limit. Therefore, if $a\to0$, the corrections associated to $\epsilon_k^r$ are vanishing.

Our simple analysis also shows that the dominant corrections are the ones associated with $\epsilon_1^t$, although in the fast-spinning case the corrections $\delta\Omega_k^r$ for different values of $k$ are all comparable to each other, and they are also comparable to $\delta\Omega_1^t$. However, at least for moderately large spin, the corrections $\delta\Omega_1^t$ and $\delta\Omega_2^r$ are the dominant ones. Note that both $\epsilon_1^t$ and $\epsilon_2^r$ are currently unconstrained by observations, so that their contribution would likely dominate the near-horizon geometry of the deformed Kerr metric~\eqref{eq:JP2_1}--\eqref{eq:JP2_5}. 
In this sense, it might be interesting to extend the detailed analysis performed using the metric~\eqref{metricJP} (see e.g. Ref.~\cite{Johannsen:2012ng,Bambi:2013eb,Bambi:2011ek,Johannsen:2013rqa,Johannsen:2013asa} and references therein) also in the case of its generalization~\eqref{eq:JP2_1}--\eqref{eq:JP2_5}, and keeping the dominant unconstrained corrections.

\subsection{Curvature singularities in the generalized JP metric}\label{sec:naked}
Similarly to what is done in Sec.~\ref{sec:naked0}, a study of the regularity of the extended metric~\eqref{eq:JP2_1}--\eqref{eq:JP2_5} is straightforward. Again, to first order in $\epsilon_k^i$
the horizon location is defined by $g^{rr}=0$ and possible singularities can only occur in the limit $J\to M_\ADM^2$, whereas the exterior metric is guaranteed to be regular everywhere if $J<M_\ADM^2$.

To simplify the result, we consider $\gamma=\beta=1$ in Eqs.~\eqref{constr1}--\eqref{constr2} and set $\epsilon_k^i=0$ when $k>5$. In the near-extremal limit we obtain
\begin{widetext}
 \begin{eqnarray}
 r_+(\theta)&&=\frac{M_\ADM^3}{2 D (3+\cos(2\theta))^3} \left\{[3+\cos(2\theta)]^2 (7+\cos(2\theta)) \epsilon_1^t-2 \sin^2\theta \left[(3+\cos(2\theta)) (3 \epsilon_2^r+\cos(2\theta) \epsilon_2^r+2 (\epsilon_3^r+\epsilon_4^r))+4 \epsilon_5^r\right]\right\}\nn\\
 &&+ {\cal O}(D^0)+{\cal O}({\epsilon_k^i}^2)\,,
\end{eqnarray}
\end{widetext}
where $D$ was defined below Eq.~\eqref{rhJP}. Using this result, one can show that the curvature invariants are everywhere regular for $J<M_\ADM^2$ and they diverge only when $J\to M_\ADM^2$, unless the conditions $\epsilon_1^t=\epsilon_2^r=\epsilon_5^r=0$ are satisfied. In the latter case, the metric is regular everywhere for $r\geq r_+(\theta)$ for any $J\leq M_\ADM^2$.

It is worth mentioning the connection between this result and the analysis of Ref.~\cite{Hansen:2013owa}. The latter paper considered a different static seed geometry (namely a perturbative solution of a modified quadratic gravity~\cite{Mignemi:1992nt,Pani:2009wy,Yunes:2011we}) and constructed the corresponding spinning geometry through the Newman-Janis algorithm. The resulting geometry is found to be singular at $r=2M_\ADM$ even in the slowly rotating case~\cite{Hansen:2013owa}. 

On the other hand, the generalized metric we present in \eqref{static2} can be specialized to the case of static BH solutions in quadratic gravity. Within our framework, such solution is described by $\epsilon_k^i\neq0$ for $k<7$ (cf. Eqs.~(9)-(10) in Ref.~\cite{Yunes:2011we}). Therefore, the analysis above shows that the rotating solution would be singular only in the limit $J\to M_\ADM^2$.

The fact that in our case the spinning metric is regular everywhere for any $J<M_\ADM^2$, whereas in Ref.~\cite{Hansen:2013owa} a singularity was found for any $J\neq0$, is due to the fact that the seed metric used in Ref.~\cite{Hansen:2013owa} cannot be recast in our form~\eqref{static2}, i.e. the function $\bar h(r)$ is not a simple expansion in $M/r$, cf. Eqs.~(31)-(35) in Ref.~\cite{Hansen:2013owa}. The difference is that the seed metric used in Ref.~\cite{Hansen:2013owa} was obtained after expressing all metric coefficients using the physical mass $M_\ADM$, whereas the procedure we presented in Sec.~\ref{sec:generalization} assumes that the Newman-Janis transformation is applied to the ``bare'' metric written in terms of $M$. This apparent conundrum shows a further drawback of the Newman-Janis transformation: even an apparently harmless mass rescaling would change the radial dependence of the seed metric and would give rise to a totally different spinning counterpart. In the particular case at hand, the difference is crucial: while the spinning geometry obtained transforming the bare static metric is regular for $J< M_\ADM^2$ to first order in the deformations, the case obtained transforming the rescaled metric gives rise to much more severe singularities which appear for any $J\neq0$.

\subsection{Matching of generalized JP metric to modified-gravity solutions}
The first requirement on any metric-parametrization candidate is that it describes known solutions. 
It turns out that the original JP seed metric \eqref{static1} fails to describe all known (to us) static metric solutions of alternative theories other than Schwarzschild. 

The generalized metric we present in \eqref{static2} on the other hand, is a generic static metric and therefore must 
describe all metric solutions in any alternative theory, at least if these solutions are analytic functions of $r$. For example, as discussed in the previous section, the parametrization~\eqref{static2} can be specialized to the case of static BH solutions in Einstein-Dilaton-Gauss-Bonnet gravity~\cite{Mignemi:1992nt,Pani:2009wy,Yunes:2011we} in the small coupling limit. Such solution is described by $\epsilon_k^i\neq0$ for $k<7$, including a nonvanishing parameter $\epsilon_1^t$ associated to a deformation of the horizon area. As expected, all nonvanishing $\epsilon_k^i$ are proportional to the fundamental coupling constant of the modified gravity. (For this reason, the small-coupling expansion adopted in this paper is rather natural if Einstein-Dilaton-Gauss-Bonnet gravity is treated as an effective field theory~\cite{Pani:2011gy}.)

Unfortunately, both the original static metric~\eqref{static1} and its extension~\eqref{static2} are affected by a common problem, in that their spinning counterparts, Eqs.~\eqref{metricJP} and \eqref{eq:JP2_1}-\eqref{eq:JP2_5}, fail to reproduce (the few) known spinning BH geometries which emerge as solutions of some modified theories of gravity (cf. also Ref.~\cite{Hansen:2013owa} for a similar analysis). In order to show that, it is sufficient to consider the slowly-rotating limit to first order in the spin.

There are at least two alternative theories which allow for slowly-rotating BHs in closed form, namely Dynamical Chern-Simons gravity~ \cite{Yunes:2009hc,Konno:2009kg,Yagi:2012ya} and Einstein-Dilaton-Gauss-Bonnet gravity~\cite{Pani:2011gy}. Together, these metrics describe the most general deformations of the Kerr metric introduced by alternative theories with all quadratic, algebraic curvature invariants generally coupled to a single scalar field. 

It is trivial to check that even the generalized metric~\eqref{eq:JP2_1}--\eqref{eq:JP2_5} fails to reproduce these solutions as particular cases in the slowly-rotating limit. This can be shown by comparing some curvature invariant at the horizon for the metric~\eqref{eq:JP2_1}--\eqref{eq:JP2_5} with those obtained from the metrics in Refs.~\cite{Yunes:2009hc,Pani:2011gy} to first order in the spin (cf. Ref.~\cite{Joao} for an extensive discussion). Since the generalization~\eqref{eq:JP2_1}--\eqref{eq:JP2_5} includes the original parametrization~\eqref{metricJP} as a particular case, the same considerations apply also to that case.

Indeed, showing that the Newman-Janis algorithm  does not work in the case of Dynamical Chern-Simons gravity is straightforward: in that case the nonspinning metric is simply Schwarzschild and the latter is mapped directly to the Kerr metric by the Newman-Janis algorithm. Since the Kerr metric differs from the spinning BH geometry in Dynamical Chern-Simons gravity to any order in the spin, it is obvious that any attempt at reproducing it in this framework will fail. Indeed, the Newman-Janis approach is simply a solution-generating technique which is useful in GR but its limitations are well known, see e.g. Ref.~\cite{Hansen:2013owa} for a more detailed analysis of this problem.

\section{Conclusions}

Forthcoming electromagnetic~\cite{Eisenhauer:2008tg,2013arXiv1311.5564B} and gravitational-wave~\cite{Berti:2009kk,Yunes:2013dva} observations will provide access to the strong-curvature region near the horizon of massive BHs for the first time, thus allowing for null tests of the Kerr hypothesis and of the no-hair theorem of GR. Developing a framework where generic deviations from the Kerr metric can be investigated in a model-independent fashion would highly facilitate such observational tests. However, constructing this framework proved to be challenging and none of the current approaches is entirely flawless~\cite{Joao}. 


We have discussed some important limitations and extensions of a proposal~\cite{Johannsen:2011dh} that has attracted much attention recently. In particular, we have shown that all previous studies have focused on subdominant corrections, neglecting the dominant terms. Even the most general parametrization in this framework --~which we presented here~-- fails to reproduce known spinning solutions in modified gravity. 

On the other hand, we have shown that considering the parametrization as a small deformation of a Kerr geometry, possible naked singularities found in Ref.~\cite{Johannsen:2013rqa} can be avoided taking particular combinations of the deformation parameters. Even when such singularities exist, to first order in the corrections they only appear in the extreme Kerr limit, $J\to M_\ADM^2$.

Perhaps most importantly, this approach shows a severe degeneracy of the extra parameters in the near-extremal limit, due to the fact that the nonspinning metric is constructed using a weak-field expansion. This latter point is particularly relevant if precise tests using these parametrizations are to be devised. 

We have shown that highly-spinning geometries are more sensitive to higher-order multipoles of the spacetime, and this makes it problematic to use parametrizations that affect only the quadrupole moment of the Kerr geometry (like the quasi-Kerr metric~\cite{Glampedakis:2005cf,2013arXiv1311.5564B}) because putative corrections to the higher-order multipoles can be equally relevant for highly-spinning objects.

While some of the limitations discussed here might be overcome (for example one might consider only moderately-spinning geometries and small deformations for which the degeneracy problem discussed above is less severe and the metric is regular everywhere), our results suggest that the parametrizations~\eqref{metricJP}, or its extension~\eqref{eq:JP2_1}--\eqref{eq:JP2_5}, should be used with extreme caution. Tests of the no-hair theorem for the supermassive object in the galactic center~\cite{2013arXiv1311.5564B} would highly benefit from the development of a more general framework, which is still lacking.

\begin{acknowledgments}
We thank Tim Johannsen for a careful reading of the manuscript and useful correspondence, and Nico Yunes for valuable comments.
V.C. acknowledges partial financial
support provided under the European Union's FP7 ERC Starting Grant ``The dynamics of black holes:
testing the limits of Einstein's theory'' grant agreement no. DyBHo--256667.
This research was supported in part by Perimeter Institute for Theoretical Physics. 
Research at Perimeter Institute is supported by the Government of Canada through 
Industry Canada and by the Province of Ontario through the Ministry of Economic Development 
\& Innovation.
P.P. acknowledges financial support provided by the European Community 
through the Intra-European Marie Curie contract aStronGR-2011-298297.
J. R. was partially funded through FCT PTDC/FIS/098025/2008 project.
This work was supported by the NRHEP 295189 FP7-PEOPLE-2011-IRSES Grant, and by FCT-Portugal through projects
PTDC/FIS/116625/2010, CERN/FP/116341/2010, CERN/FP/123593/2011 and IF/00293/2013.
\end{acknowledgments}

\bibliography{spinning}
\end{document}